\documentclass[journal,onecolumn]{IEEEtran}

\usepackage{lineno,hyperref}
\modulolinenumbers[5]
\usepackage{cite}
\usepackage{amsmath,amssymb,amsfonts}
\usepackage{graphicx}
\usepackage{textcomp}
\usepackage{multirow}
\usepackage{caption}
\usepackage{blindtext}
\usepackage{xcolor}
\usepackage[utf8]{inputenc}
\usepackage[T1]{fontenc} 
\usepackage{mathtools}
\usepackage{tabularx}
\usepackage{subfigure}
\usepackage{dblfloatfix}
\usepackage[justification=centering]{caption}

\usepackage{multicol}
\usepackage{lipsum}
\usepackage{mwe}
\usepackage{longtable}
\usepackage[normalem]{ulem}
\usepackage{algorithm,algpseudocode}
\usepackage{varwidth}

\begin{document}

\title{Pneumonia Detection on chest X-ray images Using Ensemble of Deep Convolutional Neural Networks}
\author{Alhassan Mabrouk, Rebeca P. Díaz Redondo, Abdelghani Dahou,  Mohamed Abd Elaziz, Mohammed Kayed
\thanks{Alhassan Mabrouk (alhassanmohamed@science.bsu.edu.eg) and Abdelghani Dahou (dahou.abdghani@univ-adrar.edu.dz) are with Mathematics and Computer Science Department, Faculty of Science, Beni-Suef University, Beni Suef 62511, Egypt;}
\thanks{Rebeca P. Díaz Redondo (rebeca@det.uvigo.es) is with atlanTTic, Universidade de Vigo; Vigo, 36310, Spain.}
\thanks{Mohamed Abd Elaziz (abd\_el\_aziz\_m@yahoo.com) is with Faculty of Computer Science and Engineering, Galala University, Suez 435611, Egypt; and with Artificial Intelligence Research Center (AIRC), Ajman University, Ajman P.O. Box 346, United Arab Emirates and with Department of Mathematics, Faculty of Science, Zagazig University, Zagazig 44519, Egypt.}
\thanks{Mohammed Kayed (mskayed@gmail.com) is with Computer Science Department, Faculty of Computers and Artificial Intelligence, Beni-Suef University, Beni Suef 62511, Egypt}
}

\maketitle

\abstract{Pneumonia is a life-threatening lung infection resulting from several different viral infections. Identifying and treating pneumonia on chest X-ray images can be difficult due to its similarity to other pulmonary diseases. Thus, the existing methods for predicting pneumonia cannot attain substantial levels of accuracy. \color{black}
Therefore, this paper presents a computer-aided classification of pneumonia, coined as Ensemble Learning (EL), to simplify the diagnosis process on chest X-ray images. Our proposal is based on Convolutional Neural Network (CNN) models \color{black}, which are pre-trained CNN models that have been recently employed to enhance the performance of many medical tasks instead of training CNN models from scratch. \color{black}
We propose to use three well-known CNN pre-trained (DenseNet169, MobileNetV2 and Vision Transformer) using the ImageNet database.
\color{black}
Then, these models are trained on the chest X-ray data set using fine-tuning. Finally, the results are obtained by combining the extracted features from these three models during the experimental phase. 
\color{black}
The proposed EL approach outperforms other existing state-of-the-art methods, and it obtains an accuracy of 93.91\% and a F1-Score of 93.88\% on the testing phase.
\color{black}
}



\section{Introduction}

Virus infection has been one of the most serious threats to human health throughout history. One of the most common viral infections is pneumonia \cite{ortiz2022automatic}. Infections caused by viruses and bacteria harm the lungs \cite{ben2022fusion}. Pneumonia symptoms are common, including pain, cough, shortness of breath, etc. \color{black}Pneumonia affects approximately 7.7\% of the world's population each year. As a result, early detection is critical for such illnesses. Thus, the task of automated medical image classification has grown significantly \cite{wang2022trends}. This task aims to diagnose medical images into pre-defined classes. Recently, Deep Learning (DL) has become one of the most common and widely used methods for developing medical image classification tasks \cite{singhal2022study}. Also, DL models produced more effective performance than traditional techniques using chest X-ray images from pneumonia patients \cite{ben2022fusion, iori2022mortality}.\color{black}

The DL architectures illustrated effective predictive ability and outperformed physicians \cite{salahuddin2022transparency}.  On chest X-ray images, multiple tasks were done on DL models, including tuberculosis identification \cite{kale2022early}, tuberculosis segmentation \cite{bellens2022evaluating}, large scale recognition \cite{zhang2022large}, COVID-19 detection \cite{le2022covid, sajun2022investigating}, and Radiograph classification \cite{furtado2022deep}\color{black}. The automated classification of chest X-ray images using DL models is growing rapidly, and choosing an appropriate region of interest (ROI) on chest X-ray images was used to discover pneumonia \cite{malhotra2022deep}.
Furthermore, applying the DL modes helps to avoid problems that take a long time to solve in traditional approaches. However, these models require large volumes of well-labeled training samples. 

To solve this problem, Transfer Learning (TL) has been developed. Due to its capacity to effectively solve the shortcomings of reinforcement learning and supervised learning, TL is becoming more widespread \cite{abd2022medical, adel2022improving}. The TL has the following types: Unsupervised, inductive, transductive, and negative learning. These types have been demonstrated to be able to tackle the DL problems \cite{niu2021distant}.  To enhance accuracy, TL is essential to have the most suggestive contextual and exclusionary capacity in the feature extraction stage for many fields \cite{lee2022combination}. For example, online scraping \cite{mabrouk2021seopinion}, social media \cite{chandrasekaran2022visual}, sentiment classification \cite{mabrouk2020deep}, and medical image classification \cite{lee2022combination}. \color{black}Therefore, this paper applies TL approaches to improve diagnostic reliability and reduce time-consuming decisions for clinicians.

We have proposed a new NN, coined as EL, which is built by training CNN models using TL methods.
\color{black}
To achieve this goal, CNN's MobileNet \cite{howard2017mobilenets}, DenseNet \cite{iandola2014densenet}, and Vision Transformer \cite{dosovitskiy2020image} methods were trained to detect pneumonia in chest X-ray images. 

We have decided to use three models to generate the proposed ensemble learning method trying to combine two approaches that, separately, have obtained promising results: on the one hand, using the best CNN models for the training stage; and, on the other hand, applying a vision transformer. Regarding the former, the authors of \cite{maselli2021hierarchical} proposed to use the best two CNN models to develop an ensemble learning solution. They built a hierarchical stacked method by using the most relevant features extracted by the selected two convolution networks, and they obtained a good performance. Additionally, Vision Transformer (VIT) has recently achieved highly competitive performance for several computer vision applications \cite{han2022survey}. What is more, VIT achieves remarkable results compared to CNN while it obtains fewer computational resources for pre-training \cite{dosovitskiy2020image}. Since the existing approaches merge CNN models for creating an ensemble method without using the transformers, as in \cite{ayan2021diagnosis}, we have decided to combine both to obtain a methodology that simultaneously merges and improves each individual proposal. That is, we propose a method that joins VIT and the selection of the best CNN models for the training stage.
\color{black}
In addition, the obtained features from the selected three models were combined using a probability-based ensemble approach to achieve good classification performance. 

\color{black}
With the aforementioned in mind, we have developed a novel method to enhance the diagnosis of pneumonia. This method is based on three well-known CNN models, which significantly improve classification performance. 
\color{black}
As a result of these processes, the following contributions to the suggested method are listed:

\begin{itemize}

\item We suggested an ensemble method that use forecasts from multiple CNN models to improve the classification results.

\item Instead of training a CNN model from scratch, we looked at appropriate transfer learning and fine-tuning methods.

\item The architecture of the proposed ensemble learning method is improved by using a batch normalization layer and a dropout layer. 

\item A comprehensive analysis of the developed method is compared to different state-of-the-art approaches using a real-world data set.

\end{itemize}

\color{black}
The rest of the paper is organized as follows: Section \ref{rw} provides a review of related works. In section \ref{pm} the existing CNN models and the proposed method are presented. The pneumonia classification performance of the proposed method is given in section \ref{ex}. Lastly, the conclusion provides future scope in section \ref{c}.

\section{Related works} \label{rw}

\color{black}
Over the past decade, many researchers have automatically used deep learning to detect lung infections and diseases from chest X-ray. 
\color{black}
For example, CheXNet is a 121-layer CNN-based approach developed by Rajpurkar et al. \cite{rajpurkar2017chexnet}. This approach was trained using 100,000 chest X-ray images from 14 different diseases. The approach was also applied using 420 chest X-ray, and the results were compared with those of radiologists. Therefore, it was found that the DL-based CNN method outperformed the average performance of radiological pneumonia detection.
In \cite{stephen2019efficient}, they trained a CNN method from scratch to retrieve features from chest X-ray images to achieve excellent classifier performance and used it to detect whether or not a patient had pneumonia, in contrast to previous studies based on traditional manual features.
\color{black}
Wu et al., in \cite{wu2020predict} suggested a method based on adaptive average filtration CNN and random forest to predict pneumonia using chest X-ray images. The adaptive filtration was applied to remove noise from the chest X-ray image, improving accuracy and making it easier to identify. Then, using dropout, a CNN model with two layers is created for extracting features. Nevertheless, more preprocessing with the adaptive filter is required to enhance CNN's classification accuracy. However, there are some issues with CNN models, which require a large amount of data with labels to be trained. Furthermore, learning a CNN architecture is computationally expensive and requires advanced machines. As a result, a transfer learning (TL) approach has been proposed to solve these problems.
\color{black}

\color{black}
Recently, the TL method has become very popular, mainly because it enables the CNN model to be more efficient, reduces costs, and requires fewer inputs \cite{cheplygina2019not}. \color{black}  Ayan and Ünver \cite{ayan2019diagnosis} used the Xception and VGG16 structures to fine-tune transfer learning. The design of Xception was substantially altered with the addition of two fully linked levels and multiple-output tiers with a SoftMax activation mechanism. As per the theory, the channel's initial layer has the greatest generality potential. The previous eight layers of the VGG16 architecture have been stopped, and the fully linked levels have been altered. Similarly, the test period for each image was 16 ms for VGG16 and 20 ms for the Xception network. In \cite{chouhan2020novel}, the methods included InceptionV3, ResNet18, and GoogLeNet. Classifier results were merged using the strong majority in this method. This means that the diagnosis goes with the group with a high proportion of first-time voters. Averaging out the model's testing results, this approach took 161 milliseconds per image.
On top of that, they were able to classify chest X-ray images with great accuracy. Pneumonia may be detected using deep CNNs, per the results of this research. We use standard algorithms as a component in our approach to categorizing data to keep computation costs minimum.
\color{black}
Rahman et al. \cite{rahman2020transfer} used transfer learning techniques on ImageNet to detect pneumonia using four pre-trained CNN architectures. They used three classification strategies to classify chest radiography images. Togacar et al. \cite{tougaccar2020deep} utilized three well-known CNN models for extracting features in the pneumonia classification task. They used the same data for training each model individually and acquired 1000 features from every CNN's last fully connected layer. For this task, these features are essential, which was reduced by the minimum redundancy maximum relevance (mRMR) feature selection method. Also, the selected features were fed into machine learning (ML) classification algorithms. Mittal et al. \cite{mittal2020detecting} suggested a CapsNet architecture for diagnosing pneumonia in chest X-ray images using multi-layered capsules. Liang and Zheng in \cite{liang2020transfer} suggested a new residual network-based trained TL approach for pneumonia diagnosis. Also, the DL model used in their study had 49 convolutional layers and 2 dense layers. Their model has a 90.05\% test accuracy. However, because of the huge number of convolutional layers used, this technique had a long execution time. In addition, Octave-Like Convolutional Neural Network \cite{chen2019drop} are considered lightweight and low computational cost neural networks which can replace the vanilla convolution operation such in driver distraction detection\cite{li2021driver}, document image segmentation \cite{das2020fast}, and tumor segmentation \cite{wang2021accurate}. Compared to the vanilla convolution, the octave CNN uses multi-frequency feature representation, which decomposes the input into low and high frequencies maps (feature representations) rather than only using the high frequency. Thus, the low-frequency feature maps represent a low-resolution representation of the input, which helps decrease unnecessary redundancy and the concept spatial dimensions.
\color{black}

To address this problem, several papers have been recently published that attempt to detect pneumonia using deep CNN methods with a lower number of convolutional layers, as in \cite{mahmoudi2022deep, chen2022automatic}. 
\color{black}
For example, Liang and Zheng \cite{liang2020transfer} used a CNN approach with residual junctions and dilated convolutional methods to identify pneumonia. While selecting chest X-ray, they revealed the influence of TL on CNN's approach. Transfer learning was used by Kermany et al. \cite{kermany2018identifying} to learn a CNN method to identify pneumonia in chest X-ray images. For classifying chest X-ray as normal vs. pneumonia, Rajaraman et al. \cite{rajaraman2018visualization} developed a new CNN-based approach. They used a region of interest (ROI) that only included the lungs rather than the entire image to learn CNN architecture. 
However, these approaches are still unable to achieve a high degree of efficiency in detecting pneumonia.

To sum up, there are interesting approaches in the state-of-the-art, but we have tried to go one step further by proposing a method that combines two different techniques: using CNN models for the training stage and taking the best one for ensemble learning, and using a vision transformer (VIT), which obtain good results. Therefore, the main difference between our proposal (EL method) and the other previous approaches is that we use an ensemble method that combines three well-known CNN models, one of them the most recent vision transformer. The obtained results are promising and lightly improve the state-of-the-art performance, with a small number of layers and features.

\color{black}
\section{Methodology} \label{pm}


\subsection{Deep Convolutional Neural Networks (DCNN) models} \label{dcnn}

\color{black}
Recently, many DCNN models have been suggested, which have been shown to enhance the productivity and effectiveness of machine learning (ML) \cite{ali2022structural, mabrouk2020deep}. Moreover, the DCNN models are among the most studied DL methods due to their capability to extract features automatically, and their adjustable structures, as in \cite{adel2022improving}. 
\color{black}
Many DL algorithms, such as MobileNet \cite{howard2017mobilenets}, and DenseNet  \cite{huang2017densely}, have incorporated the concept of depth-wise separable convolutions to address the disadvantages of traditional operation. In contrast to traditional convolution operations, depth-wise separable convolutions are performed independently of each input. Consequently, the algorithms are cost-effective to run and can be trained with fewer parameters in a short time. Therefore, the ensemble method has been recently introduced to learn more complex feature representations compared to single network \cite{rahman2021approach}. 

There are two kinds of ensemble techniques utilized in CNN architectures \cite{ayan2021diagnosis}. 
In the first technique, some researchers employed different CNN algorithms to obtain features from the medical images, as in \cite{gupta2021novel}. The collected features are aggregated and used in various machine learning techniques for classification/categorization tasks. Two distinct training methods and sophisticated algorithms are some of the limitations of this technique. In the second technique, predicted values are merged using a computational formula, as suggested in \cite{kassani2019classification}. The benefit of this technique is that the ensemble method correctly classifies the data due to the resulting performed by other CNN models' correct predictions. Therefore, this paper employs an ensemble technique to improve the performance of the classification task. 

\subsubsection{MobileNet}

The MobileNet architecture was designed by Howard et al. \cite{howard2017mobilenets}. The MobileNet design is based on separable convolution layers and consists of two components:
(a) Depthwise convolution: A single filter is applied to each input channel. (b) Pointwise convolution: a $1 \times 1$ convolution aggregates the depthwise convolution's outcomes. In a typical convolution, we filter and aggregate input images into a new vector of features through one phase.

Depthwise convolution is used to cut down on calculation time and model size. Eventually, MobileNet employs batch normalization and ReLU as a non-linearity activation function. Furthermore, before the fully-connected layer, the last average pooling decreases the spatial resolution to just one. 

\subsubsection{DenseNet}

The DenseNet was suggested by Huang et al. \cite{huang2017densely} for improving the depth of CNN. This approach was first implemented to address issues when CNNs became more complex in the model size. 
The authors solved the issue by linking each layer completely to the next, thus assuring maximal information and gradient transfer. One of the key benefits of adopting such a structure is that the DenseNet structure maximizes its capacity by reducing the usage of a deep or broad design via feature recycling.
Unlike traditional CNNs, DenseNet does not train duplicate features. Thus it requires fewer parameters. Moreover, since the structure has relatively thin layers, it only adds a tiny number of new feature maps. Also, the structure depends on each layer having immediate access to the gradients from the loss function and the input image during the training stage.

It is worth noting that the DenseNet concatenates the layer's return image features with the input feature maps, and thus there is no aggregate between them. The feature maps could be the same dimension to accomplish this combination in any instance. To overcome this issue, DenseBlocks are a concept introduced by DenseNet. DenseBlocks are used to ensure that the size of feature maps stays consistent inside a block while the number of filters varies among them. Layers of a particular sort (called transition layers) are put in the DenseBlocks. Also, down sampling is performed using batch normalization, a 1x1 convolution, and 2x2 layers in these layers.

\subsubsection{Vision Transformer (VIT)}

\color{black}
The VIT has successfully obtained perfect performance on different computer vision tasks, as discussed in \cite{han2022survey}. 
\color{black}
The Vision Transformer (VIT) \cite{dosovitskiy2020image} divides an image into patches and uses a transformer to pattern the similarity among these patches as sequences, resulting in sufficient image classification performance. VIT's structure can be summarized as follows: 1) Divide the given image into patches. 2) Flatten patches and use these patches to produce lower-dimensional linear embeddings (Patch Embedding). 3) Add a class token and positional embedding. 4) Give the patch sequence into the transformer layer and use a class token to get the label. 5) To get the output prediction, transfer the class token values to the Multi-Layer Perception (MLP).

Regarding inserting a 112 x 112 image to generate the patch, we start with 16 x 16 non-overlapping and overlapping patches. Therefore, generating 49 patches becomes easy and inserting them into the linear projection layer. Taking into account that each patch has three color channels. Also, the patches are loaded into the linear projection layer to achieve a long vector representation of each patch.

The overall number of overlapping and non-overlapping patches in the patch embedding is 49, and the patch size with the number of channels is 16 x 16 x 3. As a result, each patch's long vector is 768, and the patch embedding matrix is 49x196. In addition, class tokens and position embedding have been added to the sequence of embedded patches. If positional encoding is not used, the transformer will not be able to retain the information. Because of the additional class token, patch embeddings are still sized 50. Lastly, the acquired representations of the class token are obtained by feeding patch embeddings with a positional encoding and a class token into the transformer layer. As a result, the transformer encoding layer produces 1x768, which is then transferred to the MLP block to give an accurate prediction.

The transformer encoder, which contains the Multi-Head Self-Attention (MHSA) block and the MLP block, is the most important element in the VIT structure. The encoded layer has 50 x 768 as input, which this layer merged into patch embeddings, positional embeddings, and class tokens. In the VIT architecture, the previous layer's inputs and outputs for the 12 layers are 50 x 768. Furthermore, the normalization layer normalizes the inputs before they are fed into the Multi-Headed Attention (MHA) block. To obtain the query, key, and value matrix in MHA, the input data is adapted into a 50x2304 (768x3) shape using a linear layer. Then, reshape these matrices into 50x3x768, where each one is represented as 50 x 768. These matrices are then reshaped once more to 12 x 50 x 64. Once these matrices are obtained, the attention process for the MHA block is performed using the following equation:

\begin{equation}
    Attention(Query,Key,Value)=softmax(\frac{Query.Key^{T}}{\sqrt{d_{Key}}}).Value
\end{equation}

\color{black}

The outputs from the MHSA block are delivered as an input to the skip connection. Then, the outcome of the skip connection is sent to the normalization layer before being delivered to the MLP block for processing. Due to significant advancements in VIT, MLP includes a local mechanism to understand local features \cite{li2021localvit}. Furthermore, depth-wise convolution is integrated into the MLP block during the first fully connected layer to reduce parameters and achieve better results. The output of the MLP block must eventually feed the skip connection to achieve the encoder layer's output.
\color{black}

\color{black}
In this paper, the vision transformer is used because it focuses on each independent patch of the image, as well as their relationships with other patches. In contrast, the convolutional network does not have this property because it uses convolutional filters to learn image features.
\color{black}

\subsection{Proposed EL method}

\begin{figure}
    \centering
    \includegraphics[width=12.5cm]{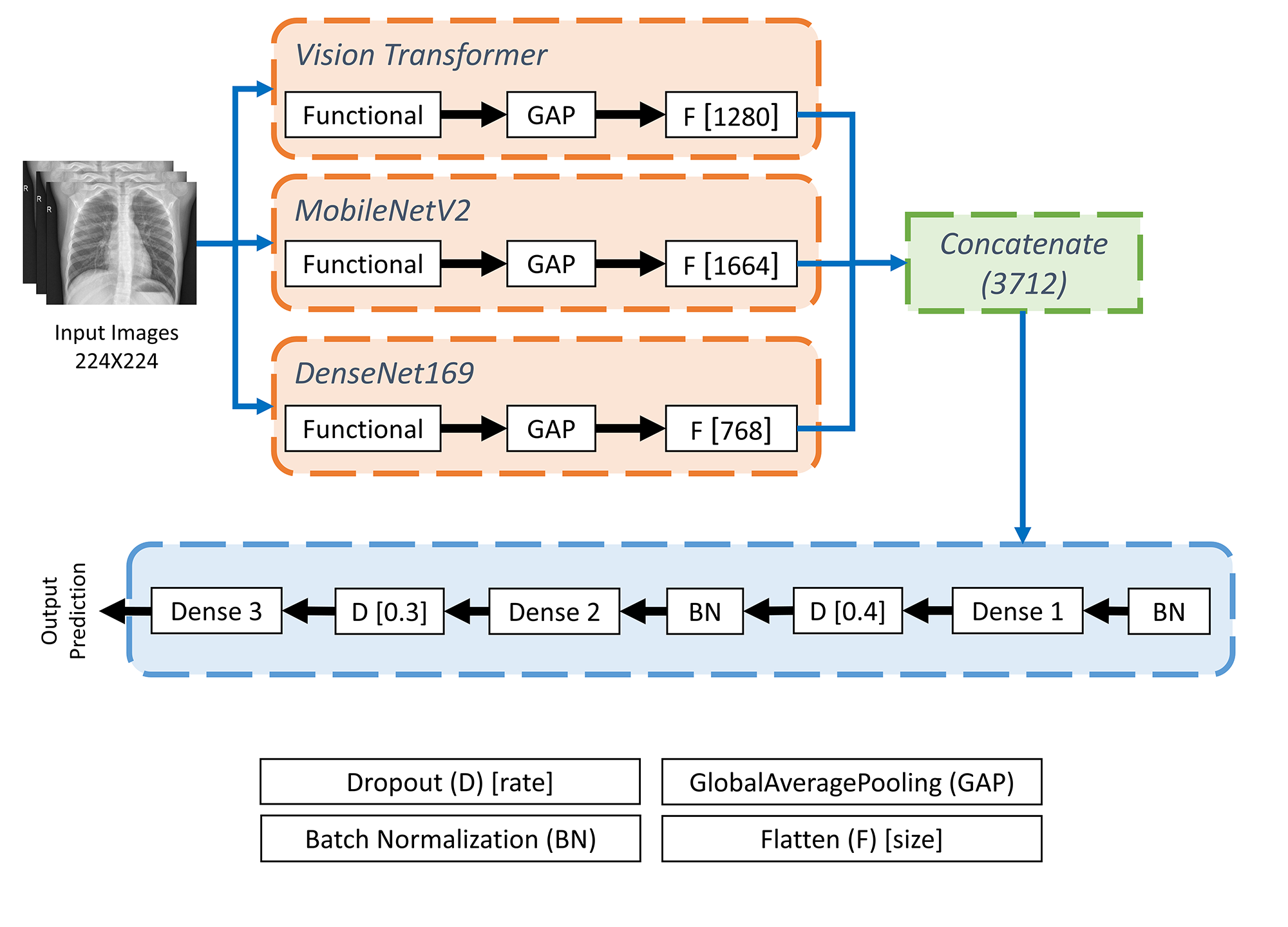}
    \caption{\label{fig:Ensemble}  The structure of the proposed ensemble learning method.}
\end{figure}

\color{black}
This section describes the implemented DL architecture based on the ensemble learning technique. The objective of the proposed method is to learn and extract medical image representation using three well-known DL models, including MobileNetV2, DenseNet169, and Vision Transformer (VIT). As shown in Figure \ref{fig:Ensemble}, the input image to the ensemble method is fed to three functional layers simultaneously. At this stage, each functional layer represents a pre-trained model that relies on MobileNetV2, DenseNet169, and VIT, respectively. For dimensionality reduction, each functional layer's output (learned representations) is fed to the global average pooling layer. After applying the pooling operation on each parallel flow, the output is flattened and concatenated to generate a single feature vector of each inputted image. To fine-tune the overall network, overcome over-fitting, and boost the classification accuracy, a sequential set of layers were placed on top of each other, including batch normalization (BN), fully-connected layer (dense), and dropout layer, as shown in Figure \ref{fig:Ensemble}. \color{black}The final output of the ensemble method is generated using a fully-connected layer to output the classification result\color{black}. 

\color{black}Using chest X-ray image data sets, the ensemble method was fine-tuned to learn and extract feature vectors from input images of size $224 \times 224$\color{black}.\color{black} These three models such as MobileNetV2, DenseNet169, and VIT were pre-trained on the ImageNet \cite{he2016deep}. In our experiments, the pre-trained ensemble method was employed and fine-tuned on the data sets having chest X-ray images. As an output, these models generate a feature vector after flattening of size 1280, 1664, and 768, respectively. Thus, the concatenated feature vector is of size 3712. During the fine-tuning of the ensemble method, the weights of the three models were fixed to accelerate the training process.

\color{black}

\section{Experimental study} \label{ex}

\color{black}
This study trained nine well-known CNN methods and the proposed EL method to classify pneumonia in chest X-ray images. In the training phase, different TL and fine-tuning techniques were attempted on these methods, and configurations ensuring excellent outcomes were utilized in the testing stage. A batch size of 32 and a learning rate of $1e-4$ were defined during this phase. We used various epoch sizes to train methods, but after 20 epochs, the methods began to overfit. To avoid overfitting of methods, early stopping was used. In addition, the Adam optimizer was applied to reduce the categorical cross-entropy loss function. For classifying, the softmax activation function was applied in the final layer.
\color{black}

As a result, this section describes the experimental study carried out. First, the data sets and the performance measures are portrayed, and then the experimental results and a discussion of them are presented. Finally, we compare our proposed method with state-of-the-art methods.

\subsection{Data set description} \label{ed}

\begin{figure}
    \centering
    \includegraphics[width=12.5cm, height=6cm]{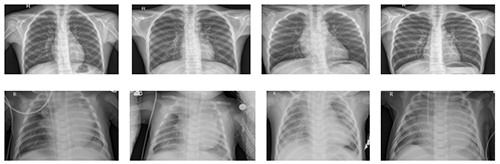}
    \caption{\label{fig:Examples}Example chest X-ray samples for classification task from the selected database.  The above row shows the normal images, and the bottom row shows the pneumonia images. }
    
\end{figure}

Pneumonia diseases have been verified for our experimental assessment, including the diagnosis of pneumonia from chest X-ray images. Figure \ref{fig:Examples}, for instance, displays a set of images from the chosen database.
The data set used in this study was provided by Kermany and Goldbaum \cite{kermany2018identifying} based on a chest X-ray scan database from pediatric patients from one to five years of age at the Guangzhou Women and Children’s Medical Center.  
In the chest X-ray Images (Pneumonia) data set, which is publicly available at \url{https://www.kaggle.com/paultimothymooney/chest-xray-pneumonia}, there are in total 5,856 normal and pneumonia chest X-ray images. 
\color{black}
To provide a fair comparison between our proposed method and the different methods. Also, the training set, the validation set, and the test set were previously divided.
The chest X-ray database was divided into two classes (normal and pneumonia). This data set contains two subsets for each class. The training subset consists of 1,341 normal patients and 3,875 pneumonia patients. Moreover, it contains 234 patients as normal and 390 pneumonia patients for the test subset. This data also consists of 16 validation data images, including eight pneumonia patients and eight normal patients. \color{black}\color{black} Examples of normal and pneumonia samples can be seen in Figure \ref{fig:Examples}\color{black}. 




\subsection{Evaluation metrics} \label{em}

\begin{equation}\label{eq_p}
    Precision = \frac{TP}{TP+FP}
\end{equation}

\begin{equation}\label{eq_re}
    Recall = \frac{TP}{TP+FN}
\end{equation}

\begin{equation} \label{eq_f}
    F1-Score = \frac{2\times Precision \times Recall}{Precision+Recall}    
\end{equation}

\begin{equation}\label{eq_ac}
      Accuracy = \frac{ TP+TN}{TP+TN+FP+FN} 
\end{equation}


\color{black}
The performance of the proposed classification method was evaluated based on precision, recall, f1-score, and accuracy, as introduced in Eqs. \eqref{eq_p}, \eqref{eq_re},  \eqref{eq_f}, and \eqref{eq_ac}, respectively. These metrics are the most popular in medical image classification \cite{kassani2019classification, ayan2021diagnosis}. The precision is measured as the percentage of exact data that conforms to specified characteristics. The recall is measured as the percentage of actual statistics to quantities that should have been explicitly anticipated. The F1-score is an indication of imbalanced data between Recall and Precision. The amount produced across all expected amounts is known as accuracy. 

According to the table, the positive term denotes pneumonia, while the negative term denotes normal images. The true term denotes the proper classification, while the false term denotes the wrong classification. The number of normal images wrongly labeled as pneumonia is called False Positive ($FP$). The number of normal images accurately recognized as normal is referred to as True Negative ($TN$). The number of normal images wrongly labeled as pneumonia is known as False Negative ($FN$). The percentage of labels found by the system is measured by the recall. The percentage of labels correctly assigned by the system is measured by precision. For providing the correct results, the F1-score is dependent on precision and recall. From a different perspective, an accuracy metric is used to evaluate the baselines for each task in the two main phases. The system's recognition rate is defined by accuracy.
\color{black}




\subsection{Results and analysis}

Initially, we chose the pre-trained models from the previous research, as in \cite{ayan2021diagnosis, perez2021ensemble}. The best three methods were selected after comparing their accuracy to other methods using the previously mentioned free dataset of chest X-ray images. In terms of testing accuracy, the MobileNetV2, VIT, and DenseNet169 models performed best, as shown in Table \ref{tab:resCom}. Their characteristics are described above in subsection \ref{dcnn}, which also includes a summary of their structures.
\color{black}

\begin{table}
\centering
\caption{\label{tab:resCom}The results of well-known CNN models.}
\begin{tabular}{|l|c|c|c|c|}
\hline
\textbf{Model} & \textbf{Precision} & \textbf{Recall} & \textbf{F1\_score} & \textbf{Accuracy} \\ \hline

Xception & 0.7971 & 0.7676 & 0.7713 & 0.7676 \\ \hline
VGG16 & 0.8126 & 0.8103 & 0.8087 & 0.8103 \\ \hline
MobileNetV2 & 0.9003 & 0.9073 & 0.9034 & 0.9087 \\ \hline
InceptionV3 & 0.8897 & 0.8871 & 0.8854 & 0.8871 \\ \hline
ResNet50 & 0.8233 & 0.8222 & 0.8226 & 0.8222 \\ \hline
DenseNet169 & 0.9133 & 0.9009 & 0.9063 & 0.9135 \\ \hline
ResNet152V2 & 0.8702 & 0.8687 & 0.8673 & 0.8687 \\ \hline
DenseNet121 & 0.8927 & 0.8922 & 0.8911 & 0.8922 \\ \hline
VIT & 0.9245 &0.9247 & 0.9244  &0.9247 \\ \hline


\end{tabular}
\end{table}

\color{black}

\begin{figure}
		\centering
		\begin{minipage}[t]{\linewidth}
			
			\includegraphics[width=0.6\textwidth]{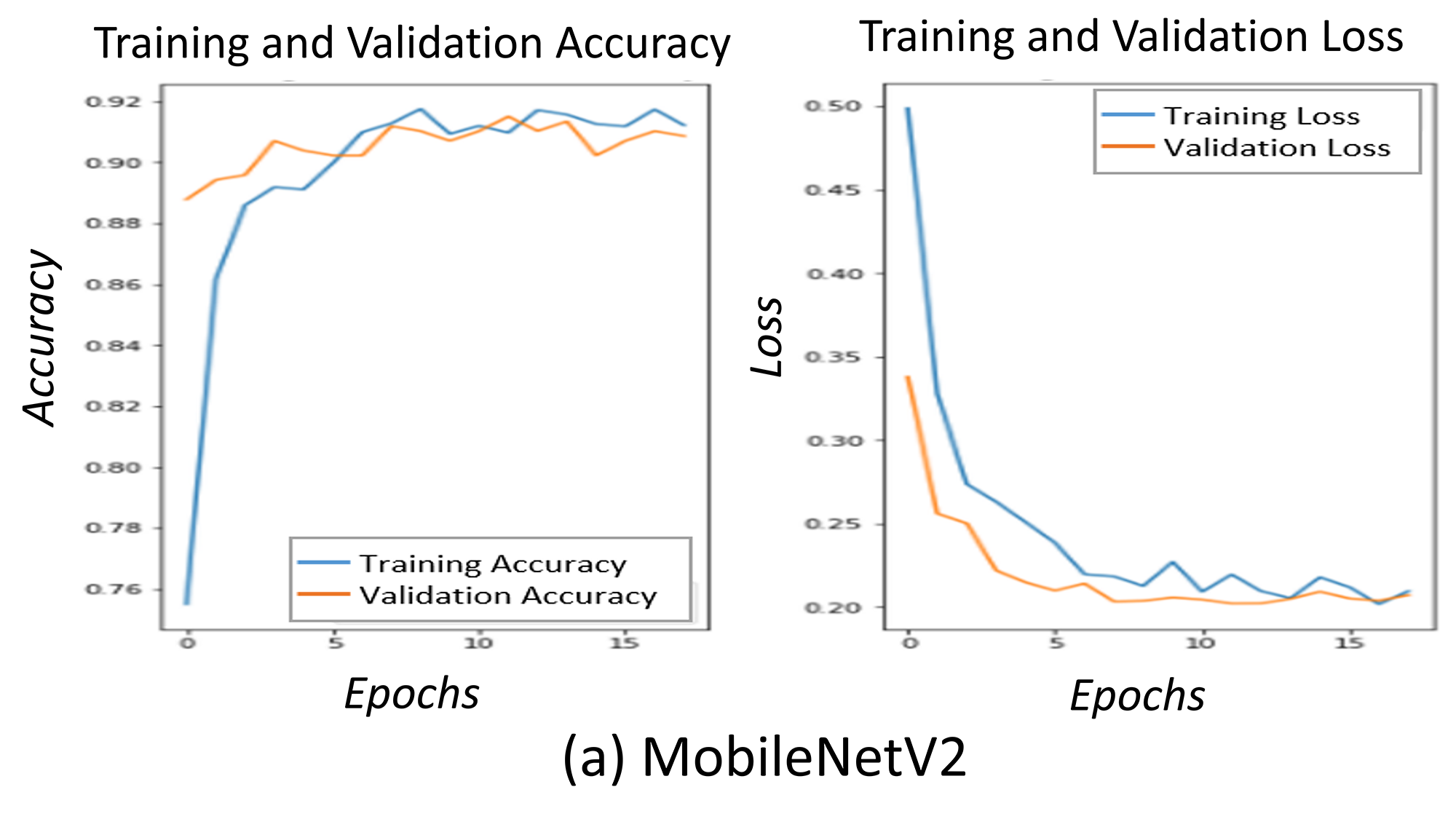}
		\end{minipage}
		\begin{minipage}[t]{\linewidth}
			
			\includegraphics[width=0.6\textwidth]{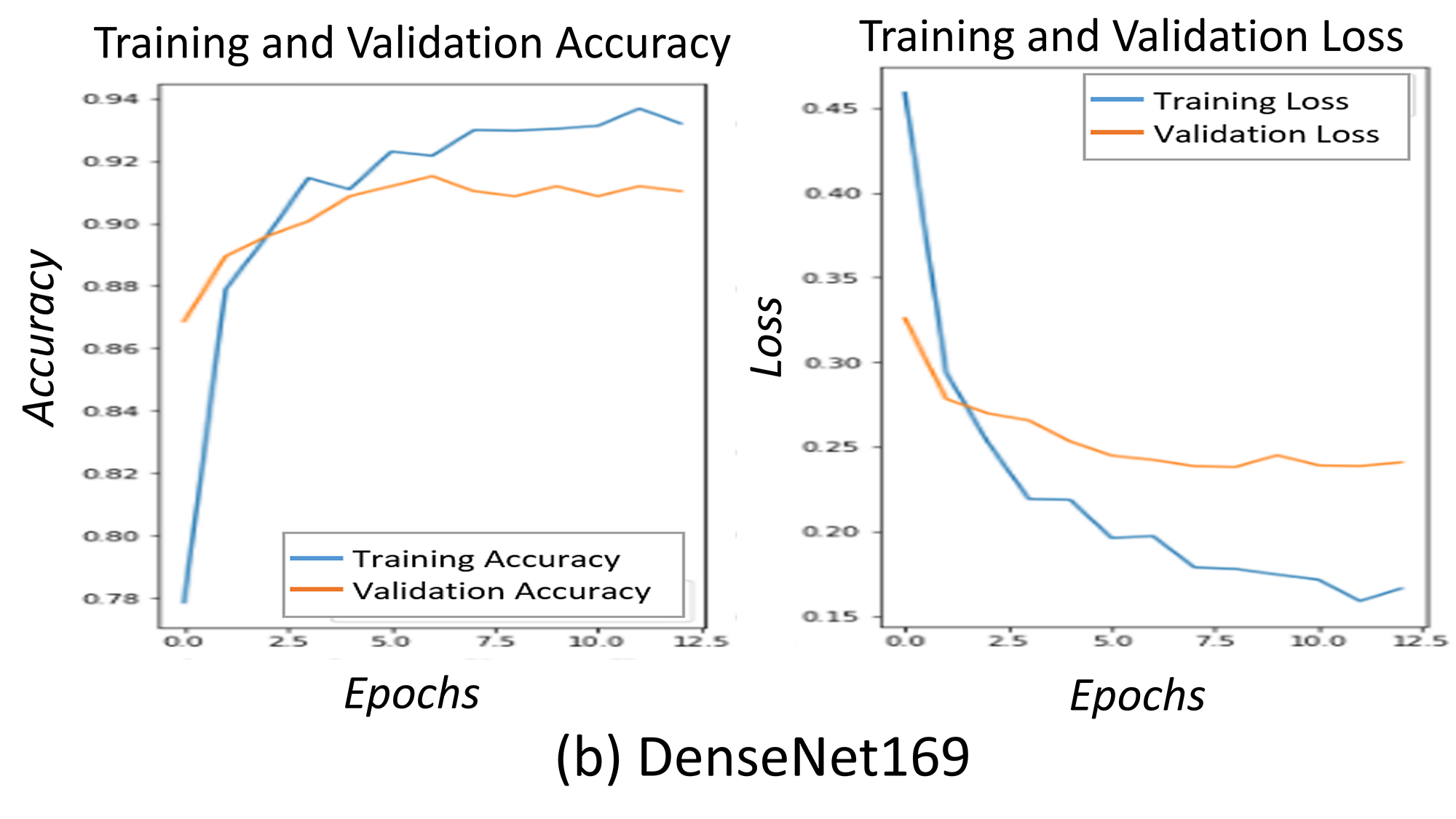}
		\end{minipage}
		\begin{minipage}[t]{\linewidth}
			
			\includegraphics[width=0.6\textwidth]{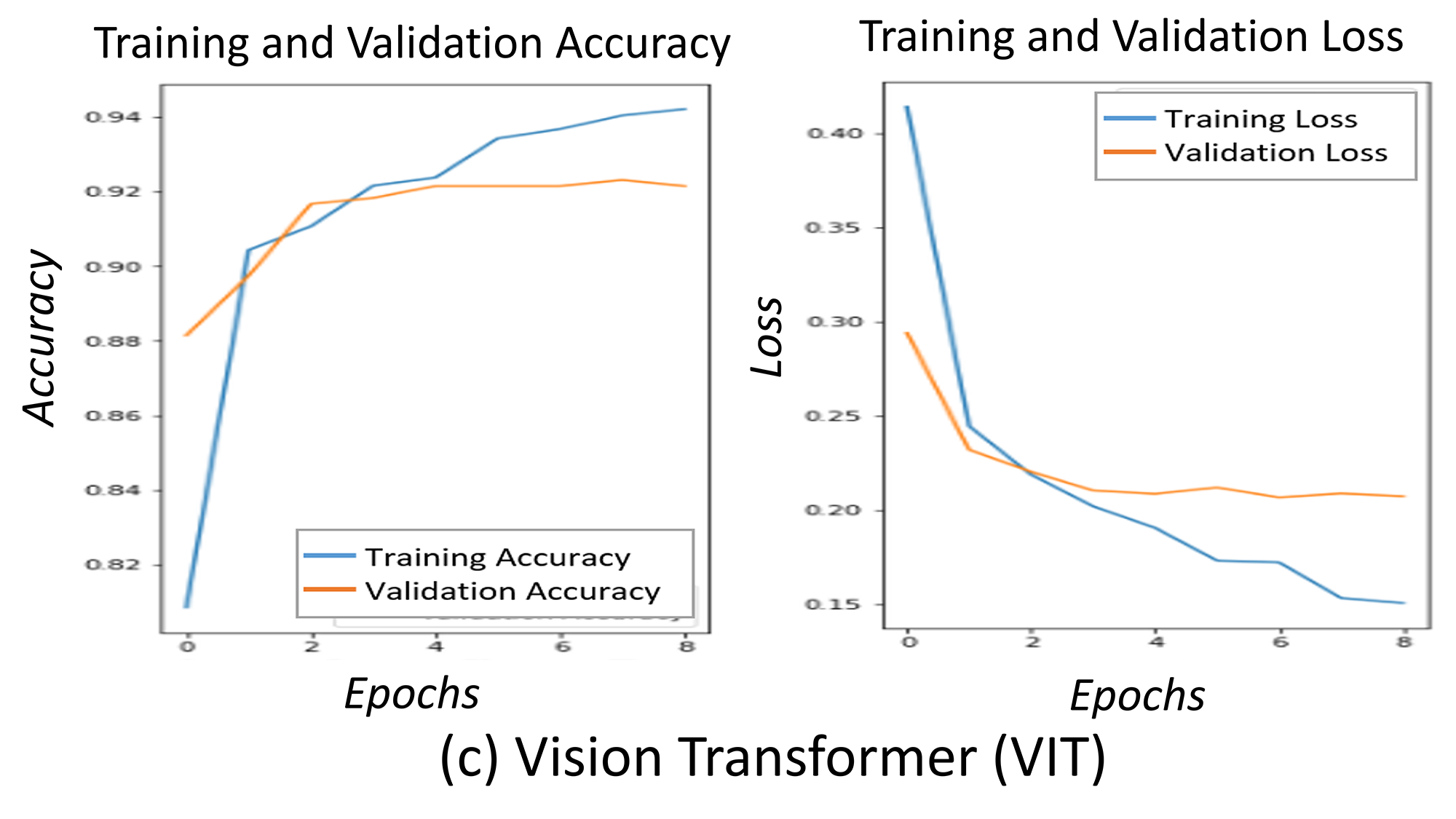}
		\end{minipage}
		\begin{minipage}[t]{\linewidth}
			
			\includegraphics[width=0.6\textwidth]{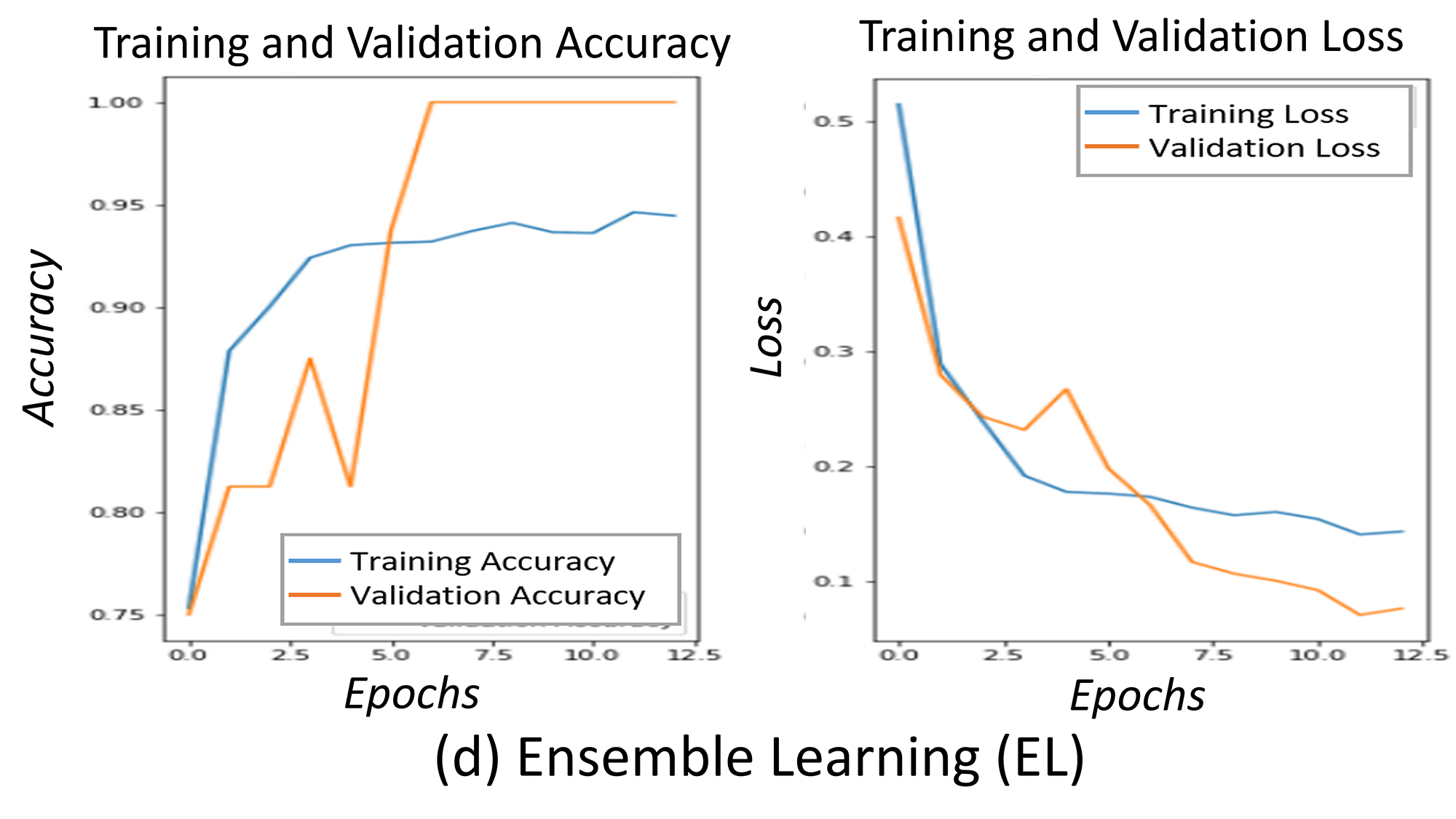}	
		\end{minipage}
		
		\caption{\label{fig:Plot}The plots referring to a) MobileNetV2, b) DenseNet169, c) Vision Transformer (VIT), and d) Ensemble Learning (EL) of the losses and accuracy on the training and validation sets.}
	\end{figure}

The performance of MobileNetV2, DenseNet169, Vision Transformer (VIT), and the proposed ensemble learning (EL) method on training and validation losses and accuracy is compared in Figure \ref{fig:Plot}. \color{black}The proposed method reduced verification loss, as shown in the figure, which improved the accuracy results\color{black}. DenseNet169's training loss is 0.1664, training accuracy is 0.9319, validation loss is 0.2408, and validation accuracy is 0.9103. Also, MobileNetV2 achieved a training accuracy of 0.9122, a training loss of 0.2096, a validation loss of 0.2072, and a validation accuracy of 0.9087. The VIT had a training loss of 0.1503, a training accuracy of 0.9421, a validation loss of 0.2071, and a validation accuracy of 0.9215. The proposed ensemble method had a training loss of 0.1361, a training accuracy of 0.9525, a validation loss of 0.0421, and a validation accuracy of 1.0.

    

The test set fulfills the requirements too. In reality, we obtain better accuracy results and a lower loss on the test set with the proposed method, illustrating a more reliable and robust approach. The accuracy of the test set can be seen in Table \ref{Res}.

\begin{table}
\centering
\caption{\color{black}Comparison of testing data results among the proposed Ensemble Learning (EL) and three well-known CNN models.\color{black}}\label{Res}

\begin{tabular}{|l|l|l|l|l|}
\hline
Model & Precision(\%) & Recall(\%) & F1-score(\%) & Accuracy(\%) \\ \hline
DenseNet169 & 91.33 & 90.09 & 90.63 & 91.35 \\ \hline
MobileNetV2 & 90.03 & 90.73 & 90.34 & 90.87 \\ \hline
VIT & 92.45 &92.47 & 92.44  &92.47 \\ \hline
\textbf{EL (Our)} & \textbf{93.96} & \textbf{92.99} & \textbf{93.43} & \textbf{93.91} \\ \hline
\end{tabular}
\end{table}

    
    
    

\begin{figure}
    \centering
    \includegraphics[width=15cm]{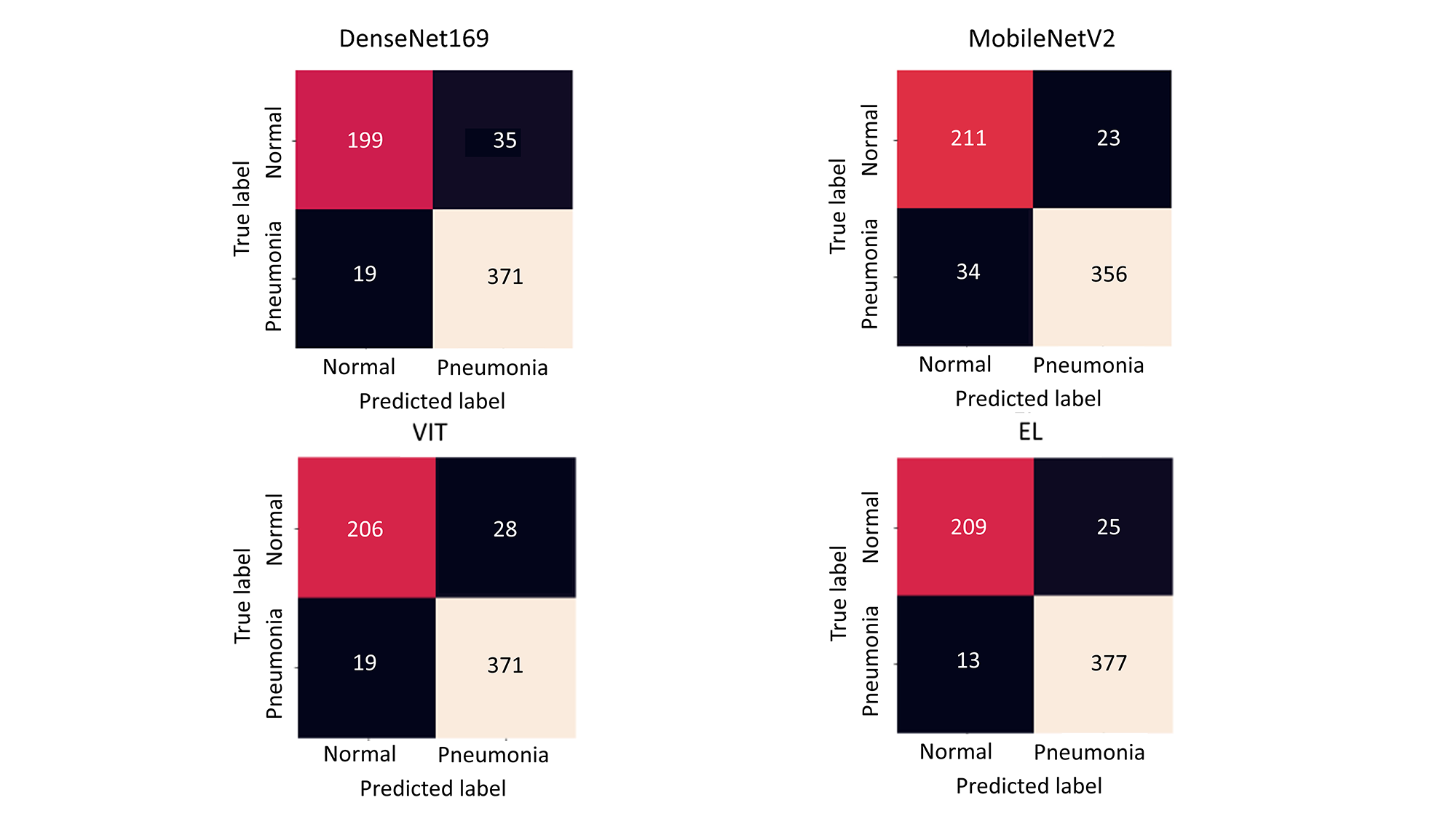}
    \caption{\label{fig:CM}  Confusion matrices of the chest X-ray data set.}
    
\end{figure}

A comparison of true and predicted labels can be seen in the confusion matrices for the proposed Ensemble Learning (EL) method, Vision Transformer (VIT), MobileNetV2 and DenseNet169 
, \color{black}as shown in Figure \ref{fig:CM}, to better understand how the four approaches did in these binary classifications. This is done primarily to understand what a good classification approach should be, as well as how it could be enhanced while trying to deal with the diagnosis of diseases, which is often critical to a patient's survival. The confusion matrices for the MobileNetV2, DenseNet169, VIT, and the proposed EL method are shown in Figure \ref{fig:CM}. The confusion matrices in the figure contain actual and predicted labels for both normal (234) and pneumonia (390) on the chest X-ray images.

\subsection{Compared Methods}\label{cm}

In this section, the proposed methodology is analyzed systematically, and its positive and negative aspects are discussed compared to other methods in the literature. The obtained results in this study were compared with other studies that achieved successful results in the literature. This
comparison is given in Table \ref{tab:resultsSOTA}. The chest X-ray data set was used to compare the various advanced methods for pneumonia detection. The state-of-the-art methods of this data set are discussed as follows: 

\begin{itemize}

\item Madani et al., \cite{madani2018chest} examined using Generative Adversarial Networks (GANs) to enrich a data set by producing chest X-ray data samples. GANs offer a method to learn about the underlying architecture of medical images, which can subsequently be used to make high-quality realistic samples. 

\item Kermany et al., \cite{kermany2018identifying} used transfer learning, which allows them to learn a neural network with a portion of the data required by traditional methods. They also made the diagnosis more transparent and understandable by highlighting the neural network's known areas.

\item Ayan and Ünver \cite{ayan2019diagnosis} employed two well-known CNN approaches, Xception and Vgg16. In the learning phase, they employed transfer learning and fine-tuning.

\item Stephen et al., \cite{stephen2019efficient} proposed a CNN-based method. Unlike other methods based solely on transfer learning or traditional handcrafted techniques, they trained the CNN model from scratch to extract attributes from a given chest X-ray image to achieve remarkable classification performance. They used it to determine if a person was infected with pneumonia or not.

\item Liang and Zheng \cite{liang2020transfer} performed pneumonia detection with a CNN model architecture using residual connections and dilated convolution methods. They also discovered the transfer learning effect on CNN models when classifying chest X-ray images.

\item Salehi et al., \cite{salehi2021automated} proposed an automatic transfer-learning method based on CNN's using DenseNet121 pre-trained concepts.

\end{itemize}

The proposed ensemble method showed better performance than a pre-trained CNN model. In addition, designing a CNN model needs massive experiments and knowledge to train a pre-trained CNN model. According to our test results, the proposed ensemble method was shown to have better performance than a pre-trained CNN model. Figure \ref{fig:CM} shows a performance comparison of the proposed MobileNetV2, DenseNet169, VIT, and the proposed EL method. In addition, designing a CNN model needs massive experiments and domain knowledge to train a pre-trained CNN model with transfer learning. Also, CNN models trained from scratch need more data, more training time, and more epochs to gain better generalization ability on input data.

However, the proposed method suffers from two drawbacks: The first is defining hyperparameters of pre-trained CNN methods while applying TL and fine-tuning to a problem of one's own. TL requires determining an appropriate pre-trained CNN method for a related issue, the size of fully connected layers, and the number of freezing layers. Many researchers use the trial-and-error approach or their own experiences to identify these parameters. Therefore, finding out TL parameters can reveal lengthy trial-and-error methods. The second drawback of the proposed EL method needs to have a lot of variance and bias.

\begin{table}
\centering
\caption{\label{tab:resultsSOTA} Comparative accuracy results for state-of-the-art method on test set of the chest X-ray data set. The best results for each item are labeled in bold.}

\begin{tabular}{|l|c|c|}
\hline
 \multicolumn{1}{|c|}{\textbf{Method/Ref.}} & \textbf{Accuracy (\%)} & \textbf{Year}\\ \hline
    DCGAN/\cite{madani2018chest}     & 84.19              & 2018\\ \hline 
    
    \cite{kermany2018identifying} & 92.80              & 2018\\ \hline 
    
    VGG16/\cite{ayan2019diagnosis}          & 87.00              &2019\\ \hline 
    
    \cite{stephen2019efficient}     & 93.73              &  2019 \\ \hline 
    
    \cite{liang2020transfer}     & 90.50              &  2020 \\ \hline 
    
    DenseNet121/\cite{salehi2021automated} & 86.80              & 2021\\ \hline 
    
    \textbf{EL}/Our & \textbf{93.91} & \textbf{2022}\\ \hline

\end{tabular}
\end{table}


\section{Conclusions} \label{c}

This paper proposes a CNN Ensemble Learning (EL) method for automatically identifying normal and pneumonia patients in chest X-ray images. For this purpose, the three most successful CNN models (DenseNet169, MobileNetV2, and Vision Transformer) were selected for the proposed EL method from among the trained CNN models. The proposed method generated the results during the testing stage. Also, a global average pooling layer was merged after the convolutional layers to avoid losing spatial information in the image. In the classifier stage of the proposed method, fully-connected layers were used. As a result, it was found that incorporating these capabilities enhances the classification performance of each CNN model. Consequently, the proposed EL method achieved satisfactory classifier performance using the chest X-ray data set. In future studies, we plan to build a weighted ensemble method based on the CNN model's accuracy.

\section{acknowledgements}
This work has received financial support from the European Regional Development Fund (ERDF) and the Galician Regional Government, under the agreement for funding the Atlantic Research Center for Information and Communication Technologies (atlanTTic). This work was also supported by the Spanish Government under re-search project “Enhancing Communication Protocols with Machine Learning while Protecting Sensitive Data (COMPROMISE)" (PID2020-113795RB-C33/AEI/10.13039/501100011033).




\bibliographystyle{ieeetr}
\bibliography{sample}


\end{document}